\documentclass{appolb}
\usepackage{psfrag}
\usepackage{graphicx}
\usepackage{amsmath}
\def\runhead{\textsc{Mueller-Navelet jets at the LHC}}

\def\slashchar#1{\setbox0=\hbox{$#1$}
   \dimen0=\wd0
   \setbox1=\hbox{/} \dimen1=\wd1
   \ifdim\dimen0>\dimen1
      \rlap{\hbox to \dimen0{\hfil/\hfil}}
      #1
   \else
      \rlap{\hbox to \dimen1{\hfil$#1$\hfil}}
      /
   \fi}

\def\bei{\begin{itemize}}
\def\ei{\end{itemize}}

\def\beeq{\begin{eqnarray}} 
\def\beqa{\begin{eqnarray}}
\def\bea{\begin{eqnarray}}

\def\eea{\end{eqnarray}}
\def\eqa{\end{eqnarray}}
\def\eeeq{\end{eqnarray}}

\def\eqar{\end{array}}
\def\beqar{\begin{array}}

\def\beas{\begin{eqnarray*}}
\def\beqas{\begin{eqnarray*}}

\def\eqas{\end{eqnarray*}}
\def\eeas{\end{eqnarray*}}

\def\beq{\begin{equation}} 
\def\be{\begin{equation}}

\def\ee{\end{equation}}
\def\eq{\end{equation}}
\def\eeq{\end{equation}}

\def\beqd{\begin{displaymath}}
\def\eeqd{\end{displaymath}}
\def\eqd{\end{displaymath}}

\def\beeq{\begin{eqnarray}} \def\eeeq{\end{eqnarray}}

\newcommand{\fin}{\end{document}}

\newcommand{\veck}{{\bf k}}

\newcommand{\veckone}{{\bf k}_1}
\newcommand{\vecktwo}{{\bf k}_2}
\newcommand{\veckj}{{\bf k}_{J}}

\newcommand{\veckjone}{{\bf k}_{J,1}}
\newcommand{\veckjtwo}{{\bf k}_{J,2}}

\newcommand{\deins}[1]{{\rm d}#1\,}
\newcommand{\dzwei}[1]{{\rm d}^2#1\,}
\newcommand{\dk}{\dzwei{\veck}}

\newcommand{\dkone}{\dzwei{\veckone}}
\newcommand{\dktwo}{\dzwei{\vecktwo}}

\newcommand{\dsigma}{\deins{\sigma}}
\newcommand{\dsigmahat}{\deins{{\hat\sigma}_{\rm{ab}}}}
\newcommand{\dnu}{\deins{\nu}}

\newcommand{\dx}{\deins{x}}
\newcommand{\dxone}{\deins{x_1}}
\newcommand{\dxtwo}{\deins{x_2}}
\newcommand{\dyjetone}{\deins{y_{J,1}}}
\newcommand{\dyjettwo}{\deins{y_{J,2}}}
\newcommand{\dphij}{\deins{\phi_{J}}}
\newcommand{\dphijone}{\deins{\phi_{J,1}}}
\newcommand{\dphijtwo}{\deins{\phi_{J,2}}}
\newcommand{\dtwojets}{{\rm d}|\veckjone|\,{\rm d}|\veckjtwo|\,\dyjetone \dyjettwo}

\newcommand{\shat}{{\hat s}}
\newcommand{\non}{\nonumber\\}
\newcommand{\asbar}{{\bar{\alpha}}_s}

\newcommand{\init}{\text{init}}
\newcommand{\MOM}{\text{MOM}}
\newcommand{\MSbar}{\overline{\text{MS}}}
\newcommand{\BLM}{\text{BLM}}

\newcommand{\avgcosn}{\langle \cos n \varphi \rangle}
\newcommand{\avgcosm}{\langle \cos m \varphi \rangle}
\newcommand{\avgcos}{\langle \cos \varphi \rangle}
\newcommand{\avgcostwo}{\langle \cos 2 \varphi \rangle}

\newcommand{\kina}{$\sqrt{s}=7$ TeV, $|\veckjone|=|\veckjtwo|=35$ GeV}
\newcommand{\kinb}{$\sqrt{s}=14$ TeV, $|\veckjone|=|\veckjtwo|=35$ GeV}
\newcommand{\kinc}{$\sqrt{s}=14$ TeV, $|\veckjone|=|\veckjtwo|=20$ GeV}
\newcommand{\kind}{$\sqrt{s}=14$ TeV, $|\veckjone|=|\veckjtwo|=10$ GeV}

\begin{document}
\title{Mueller-Navelet jets at the LHC
\thanks{Presented at the 16th conference on Elastic and Diffractive Scattering (EDS Blois 2015).}
}
\author{B. Duclou\'e$^{1,2}$, L. Szymanowski$^{3}$, S. Wallon$^{4,5}$
\address{$^{1}$Department of Physics, University of Jyv\"{a}skyl\"{a}, P.O. Box 35, 40014 University of Jyv\"{a}skyl\"{a}, Finland \\
\vspace{.2cm}
$^{2}$Helsinki Institute of Physics, P.O. Box 64, 00014 University of Helsinki, Finland \\
\vspace{.2cm}
$^{3}$National Centre for Nuclear Research (NCBJ), Warsaw, Poland \\
\vspace{.2cm}
$^{4}$Laboratoire de Physique Th\'{e}orique, UMR 8627, CNRS, Univ. Paris Sud, Universit\'{e} Paris Saclay, 91405 Orsay, France \\
\vspace{.2cm}
$^{5}$UPMC Universit\'{e} Paris 6, Facult\'{e} de physique, 4 place Jussieu, 75252 Paris Cedex 05, France
}
}
\maketitle
\begin{abstract}
We report on our NLL BFKL studies of Mueller-Navelet jets.
We first perform a complete NLL BFKL analysis supplemented by a BLM renormalization scale fixing procedure, which is successfully compared with recent CMS data. Second, we argue for the need of a measurement of an asymmetric jet configuration
in order to perform a valuable comparison with fixed order approaches. Third, we predict that the energy-momentum
violation is rather tiny in the NLL BFKL approach, for an asymmetric jet configuration. Finally, we argue that the 
double parton scattering contribution is negligible in the kinematics of actual CMS measurements.
\end{abstract}

\section{Introduction}

The high energy dynamics of
QCD, described by the Balitsky-Fadin-Kuraev-Lipatov (BFKL)
approach~\cite{Fadin:1975cb,Kuraev:1976ge,Kuraev:1977fs,Balitsky:1978ic}, have been the subject of intense studies  since four decades. The 
production of two forward jets separated by a large interval of rapidity at 
hadron colliders, as proposed by Mueller and Navelet~\cite{Mueller:1986ey}, is one of the most promising observable in order to reveal these dynamics. We here report on our study of this process in a next-to-leading logarithmic (NLL) BFKL approach.

The BFKL treatment involves two main building blocks, 
 the jet 
vertex
and the Green's function. Our complete NLL BFKL analysis of Mueller-Navelet jets, including the NLL corrections both to the Green's function~\cite{Fadin:1998py,Ciafaloni:1998gs} and to
the jet vertex~\cite{Bartels:2001ge,Bartels:2002yj},  demonstrated that the NLL corrections to the jet vertex have a very large effect, leading to a lower cross section and a much larger azimuthal correlation~\cite{Colferai:2010wu}. However,  these findings are very dependent on the choice
of the scales, especially the renormalization scale $\mu_R$ and the
factorization scale $\mu_F$, a fact which remains true when using realistic kinematical cuts for LHC experiments~\cite{Ducloue:2013hia}. In order to reduce this dependency, we then used the Brodsky, Lepage and Mackenzie (BLM) scheme~\cite{Brodsky:1982gc}. 
The net result is that one can obtain a very satisfactory description~\cite{Ducloue:2013bva}
of the most recent LHC data extracted by the CMS collaboration for the azimuthal correlations of these jets~\cite{CMS-PAS-FSQ-12-002,Safronov}.

After a recall of these NLL results, we discuss the relevance of energy-momentum conservation in our NLL BFKL treatment.
We then evaluate the importance of the potential contribution of multiparton interaction (MPI).

\section{BFKL approach}

The production of two jets of transverse momenta $\veckjone$, $\veckjtwo$ and rapidities
$y_{J,1}$, $y_{J,2}$ is described by the differential cross-section 
\beqa
\label{collinear}
  \frac{\dsigma}{\dtwojets} &=& \sum_{{\rm a},{\rm b}} \int_0^1 \dxone \int_0^1 \dxtwo f_{\rm a}(x_1) f_{\rm b}(x_2) \nonumber \\
  &\times& \frac{\dsigmahat}{\dtwojets},
\eqa
where $f_{\rm a, b}$ are the usual collinear partonic distributions (PDF). In the BFKL framework, the partonic cross-section reads
\beqa
  &&\frac{\dsigmahat}{\dtwojets} \nonumber \\
  &=& \int \dphijone\dphijtwo\int\dkone\dktwo V_{\rm a}(-\veckone,x_1)\,G(\veckone,\vecktwo,\shat)\,V_{\rm b}(\vecktwo,x_2),\label{eq:bfklpartonic}
\eqa
where $V_{\rm a, b}$ and $G$ are respectively the jet vertices and the BFKL Green's function. One should note that the use of conventional PDF in Eq.~(\ref{collinear}) is justified by the fact that the rapidity $Y=y_{J,1}-y_{J,2}$ is large enough so that the momentum fractions $x_1$ and $x_2$ are not parametrically small.
Besides the cross section, the azimuthal correlation of the two jets is another relevant observable sensitive to resummation effects~\cite{DelDuca:1993mn,Stirling:1994he}. Denoting as $\phi_{J,1}$, $\phi_{J,2}$ the azimuthal angles of the two jets, and defining
the relative azimuthal angle $\varphi$ such that $\varphi=0$ corresponds
to the back-to-back configuration, the moments of this distribution read
\begin{equation}
  \langle\cos(n\varphi)\rangle \equiv \langle\cos\big(n(\phi_{J,1}-\phi_{J,2}-\pi)\big)\rangle = \frac{\mathcal{C}_n}{\mathcal{C}_0} \,,
\end{equation}
with
\begin{equation}   
\mathcal{C}_0 = \frac{\dsigma}{\dtwojets} \,,
\end{equation}
and
\begin{equation}
  \mathcal{C}_n = (4-3\delta_{n,0}) \int \dnu C_{n,\nu}(|\veckjone|,x_{J,1})C^*_{n,\nu}(|\veckjtwo|,x_{J,2}) \left( \frac{\shat}{s_0} \right)^{\omega(n,\nu)}\,.
  \label{Cn}
\end{equation}
The coefficients $C_{n,\nu}$ are given by
\begin{equation}
   C_{n,\nu}(|\veckj|,x_{J})= \int\dphij\dk \dx f(x) V(\veck,x) E_{n,\nu}(\veck) \cos(n\phi_J)\,,
  \label{Cnnu}
\end{equation}
where 
\begin{equation}
  E_{n,\nu}(\veck) = \frac{1}{\pi\sqrt{2}}\left(\veck^2\right)^{i\nu-\frac{1}{2}}e^{in\phi}\,.
\label{def:eigenfunction}
\end{equation}
At leading logarithmic (LL) accuracy, the jet vertex reads
\begin{equation}
  V_{\rm a}(\veck,x)=V_{\rm a}^{(0)}(\veck,x) = \frac{\alpha_s}{\sqrt{2}}\frac{C_{A/F}}{\veck^2} \delta\left(1-\frac{x_J}{x}\right)|\veckj|\delta^{(2)}(\veck-\veckj)\,,
\end{equation}
with $C_A=N_c=3$ (incoming gluon) and $C_F=(N_c^2-1)/(2N_c)=4/3$ (incoming quark). The expressions of the next-to-leading order (NLO) corrections to $V_{\rm a}$~\cite{Bartels:2001ge,Bartels:2002yj,Caporale:2011cc,Hentschinski:2011tz,Chachamis:2012cc}, can be found in ref.~\cite{Colferai:2010wu}. They have been computed in the limit of small cone jets in ref.~\cite{Ivanov:2012ms} and used in refs.~\cite{Caporale:2012ih,Caporale:2013uva,Caporale:2014gpa,Caporale:2015uva,Celiberto:2015yba} (see also~\cite{Colferai:2015zfa}). The LL BFKL trajectory reads 
\begin{equation}
  \omega(n,\nu) = \asbar \left[ 2 \, \Psi(1)-\Psi \left(\frac{n+1}{2}+ i \nu\right)-
 \Psi \left(\frac{n+1}{2}- i \nu\right)
  \right],
\end{equation}
where $\asbar = N_c\alpha_s/\pi$, while at NLL, its analytical expression is much more involved~\cite{Kotikov:2000pm,Kotikov:2002ab,Ivanov:2005gn,Vera:2006un,Vera:2007kn,Schwennsen:2007hs}, see ref.~\cite{Ducloue:2013hia} for explicit formulas.

Even at NLL accuracy, several observables depend strongly on the choice of the scales, and in particular the renormalization scale $\mu_R$. An
optimization procedure to fix the renormalization scale allows to reduce
this dependency. We use the BLM procedure~\cite{Brodsky:1982gc}, which is a way of absorbing the non conformal
terms of the perturbative series in a redefinition of the coupling constant, to
improve the convergence of the perturbative series.
The first practical implementation of the BLM procedure in the context of BFKL was performed in refs.~\cite{Brodsky:1996sg,Brodsky:1997sd,Brodsky:1998kn,Brodsky:2002ka}. In refs.~\cite{Brodsky:1998kn,Brodsky:2002ka} it was argued that, when dealing with BFKL calculations, the BLM procedure is more conveniently applied in a physical renormalization scheme like the $\MOM$ scheme instead of the usual $\MSbar$ scheme, a method followed in
refs.~\cite{Angioni:2011wj,Hentschinski:2012kr,Hentschinski:2013id}.
The observables introduced above in the $\MSbar$ scheme can be obtained in the  $\MOM$ scheme using~\cite{Celmaster:1979dm,Celmaster:1979km}
\begin{equation}
  \alpha_{\MSbar}=\alpha_{\MOM}\left(1+\alpha_{\MOM}\frac{T_{\MOM}}{\pi}\right)\,,
\end{equation}
where $T_{\MOM}=T_{\MOM}^\beta+T_{\MOM}^{conf}$,
\begin{eqnarray}
   T_{\MOM}^{conf} &=& \frac{N_c}{8}\left[\frac{17}{2}I+\frac{3}{2}\left(I-1\right)\xi+
   \left(1-\frac{1}{3}I\right)\xi^2-\frac{1}{6}\xi^3\right], \non
   T_{\MOM}^\beta &=& -\frac{\beta_0}{2} \left(1+\frac{2}{3}I\right),
\end{eqnarray}
where  $I=-2\int_0^1 dx \ln(x)/[x^2-x+1] \simeq 2.3439$ and $\xi$ is the covariant gauges parameter.
Performing the transition from the $\MSbar$ to the $\MOM$ schemes, one should then choose the renormalization scale to make the $\beta_0$-dependent part vanish. This is achieved with
\begin{equation}
  \mu^2_{R,\BLM}=|\veckjone|\cdot|\veckjtwo| \exp \left[ \frac{1}{2}
  \chi_0(n,\gamma)-\frac{5}{3}+2\left(\!1+\frac{2}{3}I\!\right)\! \right]\!.
\end{equation}

\section{Results: symmetric and asymmetric configurations}

\subsection{Symmetric configuration and CMS data}

We first compare our results with the measurement performed by the CMS collaboration on the azimuthal correlations of Mueller-Navelet jets at the LHC at a center of mass energy $\sqrt{s}=7$ TeV~\cite{CMS-PAS-FSQ-12-002}.
The two jets have transverse momenta larger than $35$ GeV and rapidities lower than $4.7$. We use the anti-$k_t$ jet algorithm~\cite{Cacciari:2008gp} with a size parameter $R=0.5$ and the MSTW 2008~\cite{Martin:2009iq} parametrization for the PDFs.
On the plots we show the CMS data (black dots with error bars), the NLL BFKL result using the ``natural'' scale choice $\mu_R=\sqrt{|\veckjone| \cdot |\veckjtwo|}$ (solid black line) and the NLL BFKL results using the BLM scale setting (gray error band\footnote{The gray error band corresponds to  the typical theoretical uncertainty when practically implementing the BLM procedure.}). 

\begin{figure}[h]
\psfrag{BLM}[l][l][.7]{\scalebox{.9}{NLL, $\mu_R=\mu_{R,\BLM}$}}
\psfrag{NLL}[l][l][.7]{\scalebox{.9}{NLL, $\mu_R=\mu_{R,\init}$}}
\psfrag{CMS}[l][l][.7]{\scalebox{.9}{CMS data}}
\psfrag{Y}[][][1]{\scalebox{.9}{$Y$}}
  \begin{minipage}{0.48\textwidth}
    \psfrag{cos}[l][l][.8]{$\avgcos$}
\hspace{-.25cm}    
\includegraphics[width=6.5cm]{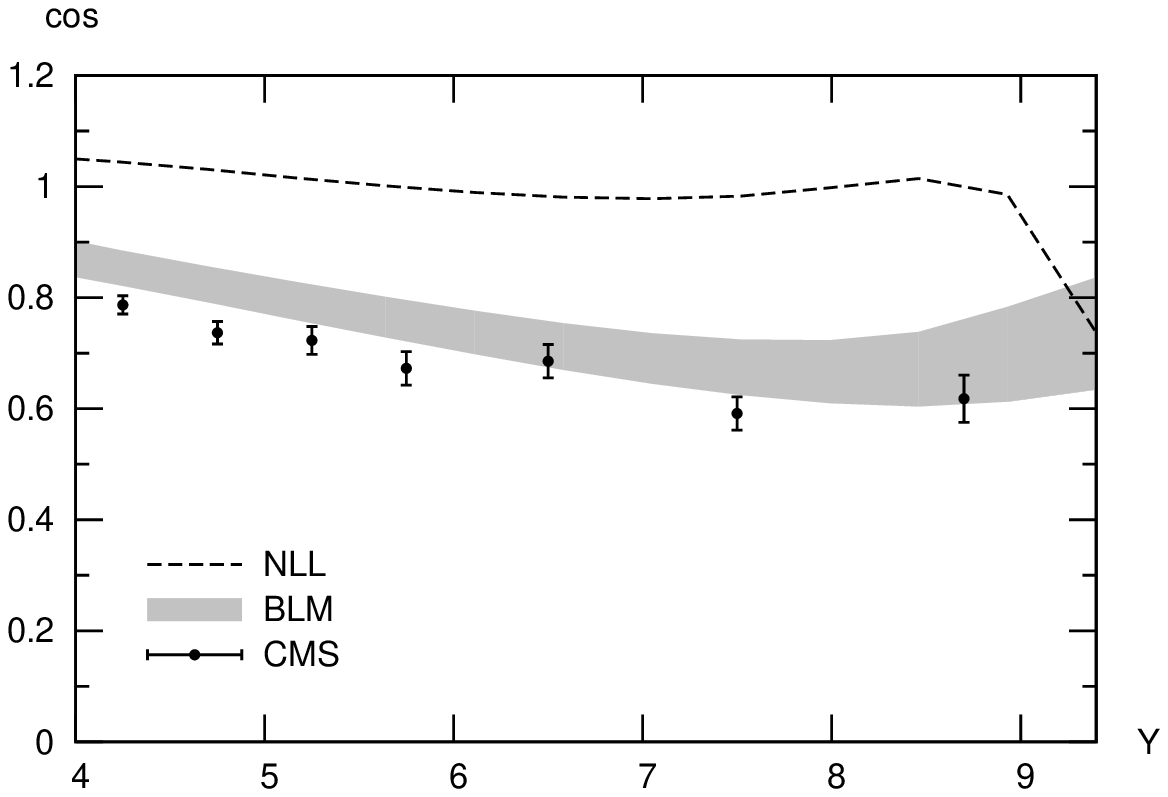}
  \end{minipage}
  \begin{minipage}{0.49\textwidth}
    \psfrag{cos}[l][l][.8]{$\avgcostwo$}
    \includegraphics[width=6.5cm]{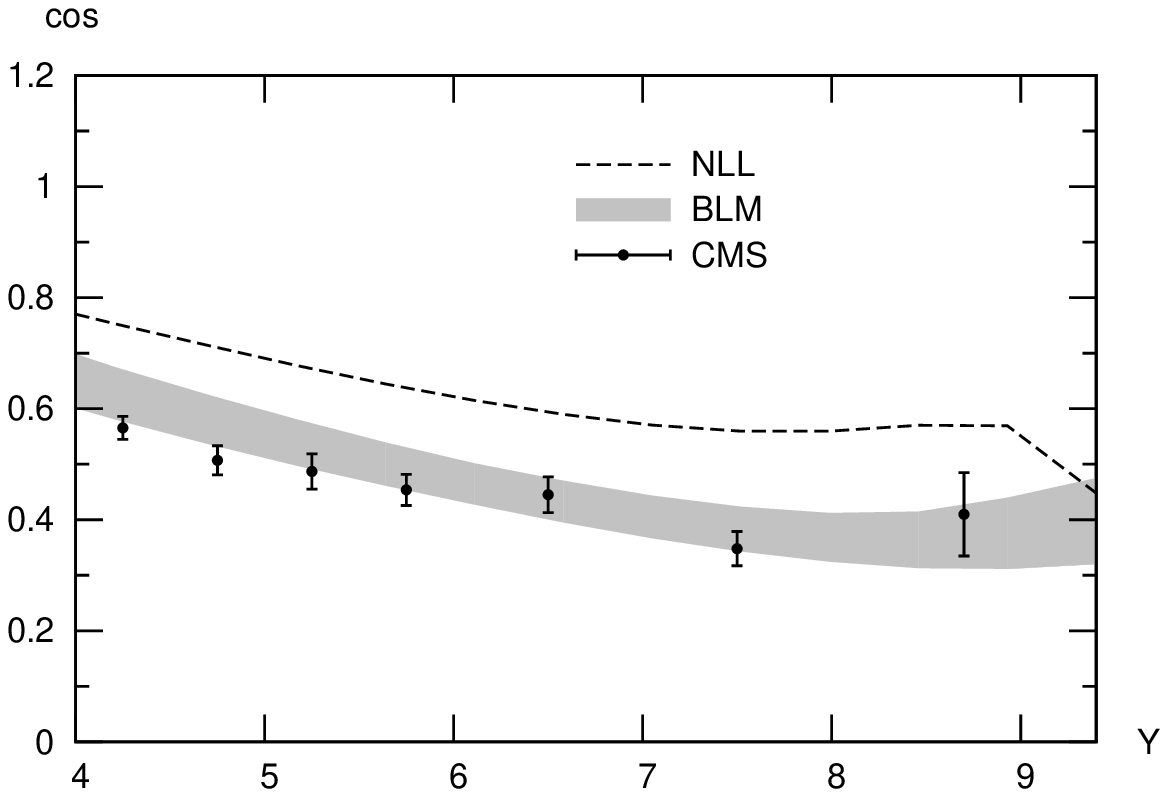}
  \end{minipage}
  \caption{Symmetric configuration. Left: Variation of $\avgcos$ as a function of $Y$ at NLL accuracy compared with CMS data. Right:
  Variation of $\avgcostwo$ as a function of $Y$ at NLL accuracy compared with CMS data.}
\label{Fig:cos-cos2_blm_sym}
\end{figure}

Our results for the angular correlations $\avgcos$ and $\avgcostwo$ as a function of $Y$ are shown in fig.~\ref{Fig:cos-cos2_blm_sym} (L) and (R) respectively.
For these two observables,  the NLL BFKL calculation with 
the ``natural'' scale choice
predicts a too strong correlation, while using the BLM procedure to fix the renormalization scale leads to a very good agreement with the data.

\begin{figure}
\psfrag{Y}[][][1]{\scalebox{.9}{$Y$}}
\psfrag{BLM}[l][l][.7]{\scalebox{.9}{NLL, $\mu_R=\mu_{R,\BLM}$}}
\psfrag{NLL}[l][l][.7]{\scalebox{.9}{NLL, $\mu_R=\mu_{R,\init}$}}
\psfrag{CMS}[l][l][.7]{\scalebox{.9}{CMS data}}
\psfrag{dist}[][][1]{$\frac{1}{\sigma}\frac{d \sigma}{d\varphi}$}
\psfrag{phi}[][][1]{\scalebox{.9}{$\varphi$}}
  \begin{minipage}{0.45\textwidth}
\hspace{-1cm}    \includegraphics[height=5cm]{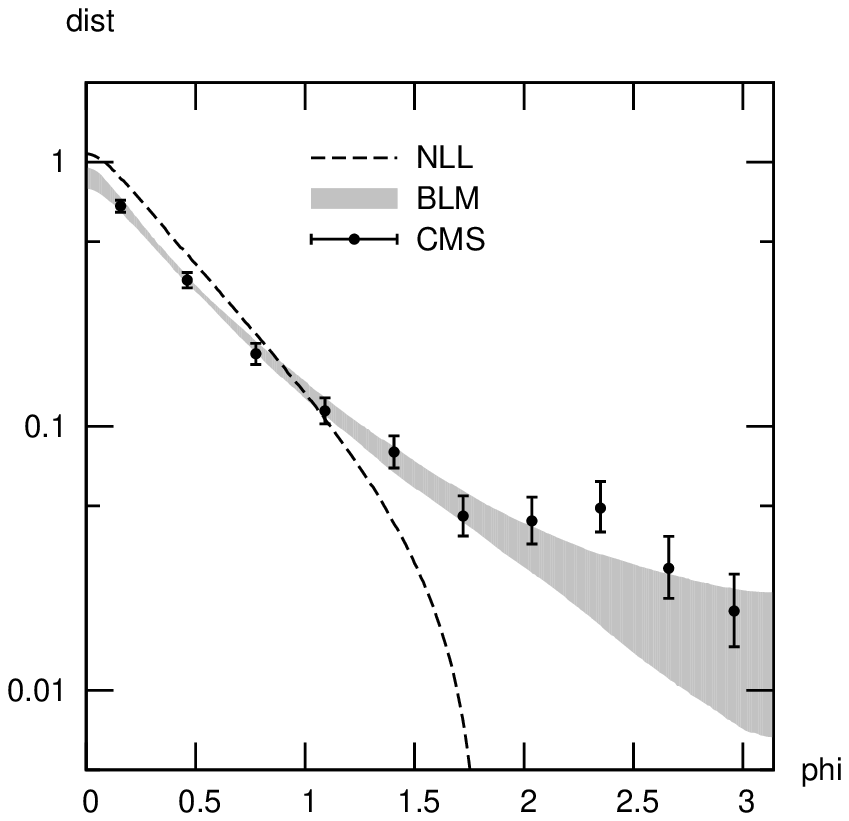}
  \end{minipage}
  \begin{minipage}{0.49\textwidth}
    \psfrag{cos}[l][l][.8]{$\avgcostwo/\avgcos$}
\hspace{-.3cm}    \includegraphics[width=7cm]{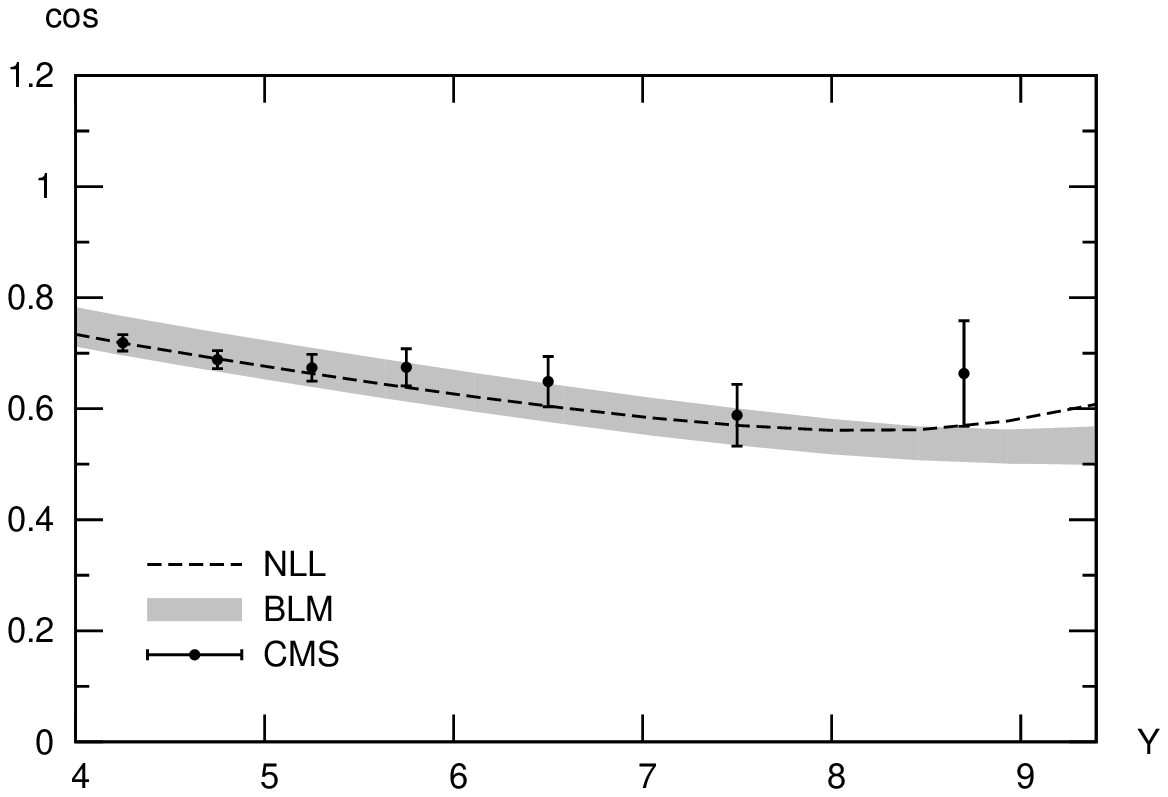}
  \end{minipage}
  \caption{Symmetric configuration. Left: Azimuthal distribution at NLL accuracy compared with CMS data.
  Right: Variation of $\avgcostwo / \avgcos$ as a function of $Y$ at NLL accuracy compared with CMS data.}
\label{Fig:cos2cos-dist_blm_sym}
\end{figure}

This improvement due to the BLM procedure is most clearly seen through 
 the azimuthal distribution of the jets $\frac{1}{{\sigma}}\frac{d{\sigma}}{d \varphi}$, 
\begin{equation}
 \frac{1}{{\sigma}}\frac{d{\sigma}}{d \varphi}
  ~=~ \frac{1}{2\pi}
  \left\{1+2 \sum_{n=1}^\infty \cos{\left(n \varphi\right)}
  \left<\cos{\left( n \varphi \right)}\right>\right\}\,,
\end{equation}
as 
displayed in fig.~\ref{Fig:cos2cos-dist_blm_sym} (L).

It was already observed both at LL and NLL accuracy~\cite{Vera:2006un,Vera:2007kn,Schwennsen:2007hs,Colferai:2010wu,Ducloue:2013hia} that ratios of the kind $\avgcosm/\avgcosn$ with $n \neq 0$ are much more stable with respect to the scales than individual moments $\avgcosn$ and therefore are almost not affected by the BLM procedure.
Indeed, fig.~\ref{Fig:cos2cos-dist_blm_sym} (R) for $\avgcostwo/\avgcos$ shows that the good agreement with the data obtained when using either the ``natural'' scale or the BLM procedure.

\subsection{Asymmetric configuration: BFKL versus fixed order}

It is well known that fixed order calculation are unstable when the lower cut on the transverse momenta of both jets is the same~\cite{Andersen:2001kta,Fontannaz:2001nq}. This is the situation encountered 
 by the above CMS measurement~\cite{CMS-PAS-FSQ-12-002}.
  Still, a comparison of the agreement of a fixed order calculation and of a BFKL one with data would be very useful to further investigate the need for resummation effects at high energy.
We now choose the lower cut on the transverse momenta of the jets to slightly differ. In practice, this is implemented by taking the same cuts as above, but now with the additional requirement that the transverse momentum of at least one jet is larger than $50$ GeV, making
the fixed order calculation now trustable.
\begin{figure}[h]
\psfrag{Y}[][][1]{\scalebox{.9}{$Y$}}
\psfrag{Dijet}[l][l][.7]{\scalebox{.9}{NLO fixed-order}}
\psfrag{BLM}[l][l][.7]{NLL BFKL, $\mu_R=\mu_{R,\BLM}$}
\psfrag{NLL}[l][l][.7]{NLL BFKL, $\mu_R=\mu_{R,\init}$}
\psfrag{cos}[l][l][.8]{$\avgcostwo/\avgcos$}
  \centering\includegraphics[width=7.5cm]{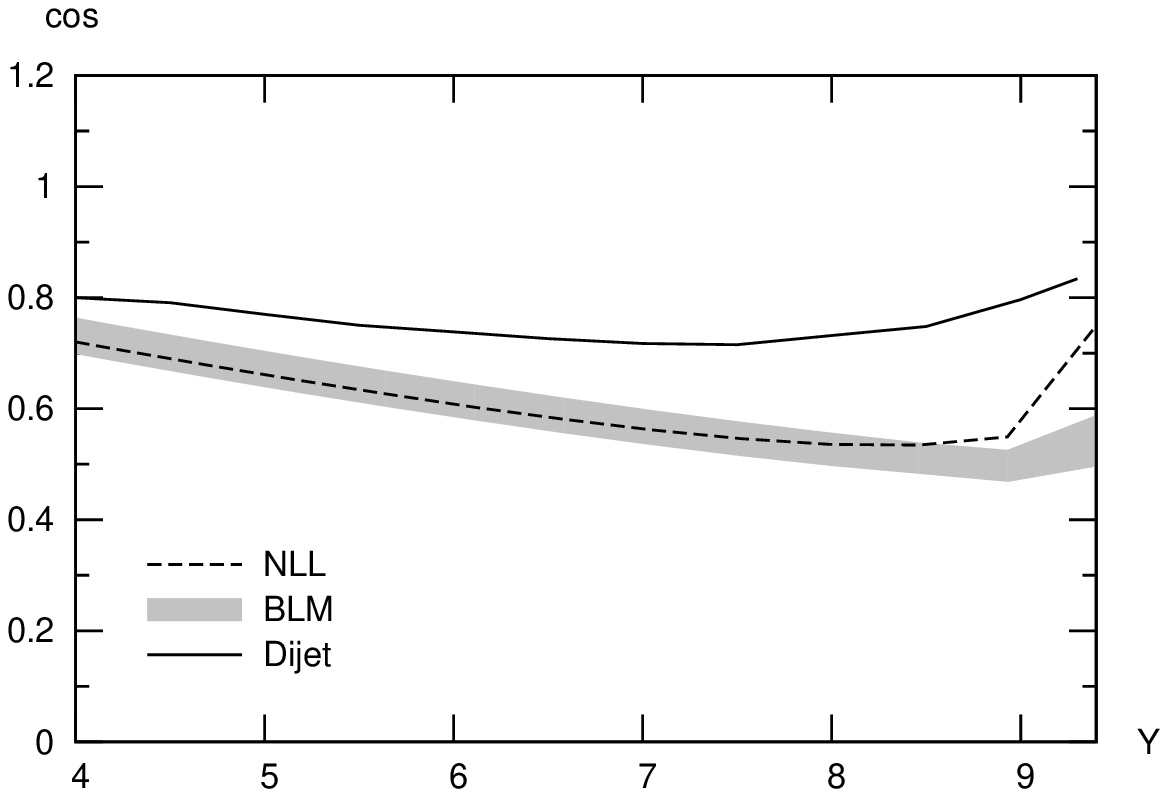}
  \caption{Asymmetric configuration. Variation of $\avgcostwo / \avgcos$ as a function of $Y$ at NLL
  accuracy compared with a fixed order treatment.}
\label{Fig:cos2cos_blm_asym}
\end{figure}
As discussed previously, the quantities $\avgcosn$ are not very stable even at NLL accuracy in the BFKL approach, so that a comparison with a fixed order calculation for these observables would not be very meaningful.
On the contrary, the observable $\avgcostwo/\avgcos$ is more stable in the BFKL approach.  Fig.~\ref{Fig:cos2cos_blm_asym} shows the comparison of the NLL BFKL calculation with the results obtained with the NLO fixed order code {\textsc{Dijet}}~\cite{Aurenche:2008dn} and clearly demonstrate that a sizable difference between the two treatments is expected over a large $Y$ range.

One should note that in the very peculiar situation where the two jets are almost back-to-back, for which a fixed order calculation is unstable, resummation effects \`a la Sudakov should be considered, to stabilize the calculation. In the BFKL approach, although this back-to-back limit is stable, the azimuthal distribution can be significantly affected by such resummation effects. These have been obtained recently in the LL approximation~\cite{us}.

\section{Energy-momentum conservation}

It is well know that energy-momentum conservation
is not satisfied in the BFKL approach,
being formally a sub-leading effect. Still, it could be numerically important, at least at LL accuracy.
It was proposed~\cite{DelDuca:1994ng} to evaluate the importance of this effect by comparing the results of an exact $\mathcal{O}(\alpha_s^3)$ calculation with the BFKL result, expanded in powers of $\alpha_s$ and truncated to order $\alpha_s^3$. This showed that a LL BFKL calculation strongly overestimates the cross section with respect to an exact calculation as long as the two jets transverse momenta are not very similar (which is the case in the asymmetric configuration discussed above). 
In the same spirit, a study with LO vertices and NLL Green's function was performed in ref.~\cite{Marquet:2007xx}. Having in mind that adding corrections beyond the LL approximation should reduce the violation of
energy-momentum conservation, we here also include NLO corrections to the jet vertices~\cite{Ducloue:2014koa}.
Consider the effective rapidity $Y_{\rm eff}$~\cite{DelDuca:1994ng}
\begin{equation}
  Y_{\rm eff} \equiv\ Y \frac{\mathcal{C}_m^{2\to3}}{\mathcal{C}_m^{{\rm BFKL},\mathcal{O}(\alpha_s^3)}} \,.
  \label{eq:Cm_e-m_cons}
\end{equation}
where $\mathcal{C}_m^{2\to3}$ is the exact $\mathcal{O}(\alpha_s^3)$ results obtained by studying the reaction $gg \to ggg$, while ${\mathcal{C}_m^{{\rm BFKL},\mathcal{O}(\alpha_s^3)}}$ is the BFKL result expanded in powers of $\alpha_s$ and truncated to order $\mathcal{O}(\alpha_s^3)$.
This effective rapidity~(\ref{eq:Cm_e-m_cons}) has the property that if one replaces $Y$ by $Y_{\rm eff}$ in the BFKL calculation, expands in powers of $\alpha_s$ and truncates to order $\alpha_s^3$, the exact result is recovered.
\psfrag{LO}[l][l][.8]{LL}
\psfrag{NLO}[l][l][.8]{NLL}
\psfrag{yeffy}[l][l][.8]{$Y_{\rm eff}/Y$}
\psfrag{kj1}[l][l][0.8]{\hspace{0.1cm}$\veckjtwo$ (GeV)}
\begin{figure}[h]
\centering\includegraphics[width=7.5cm]{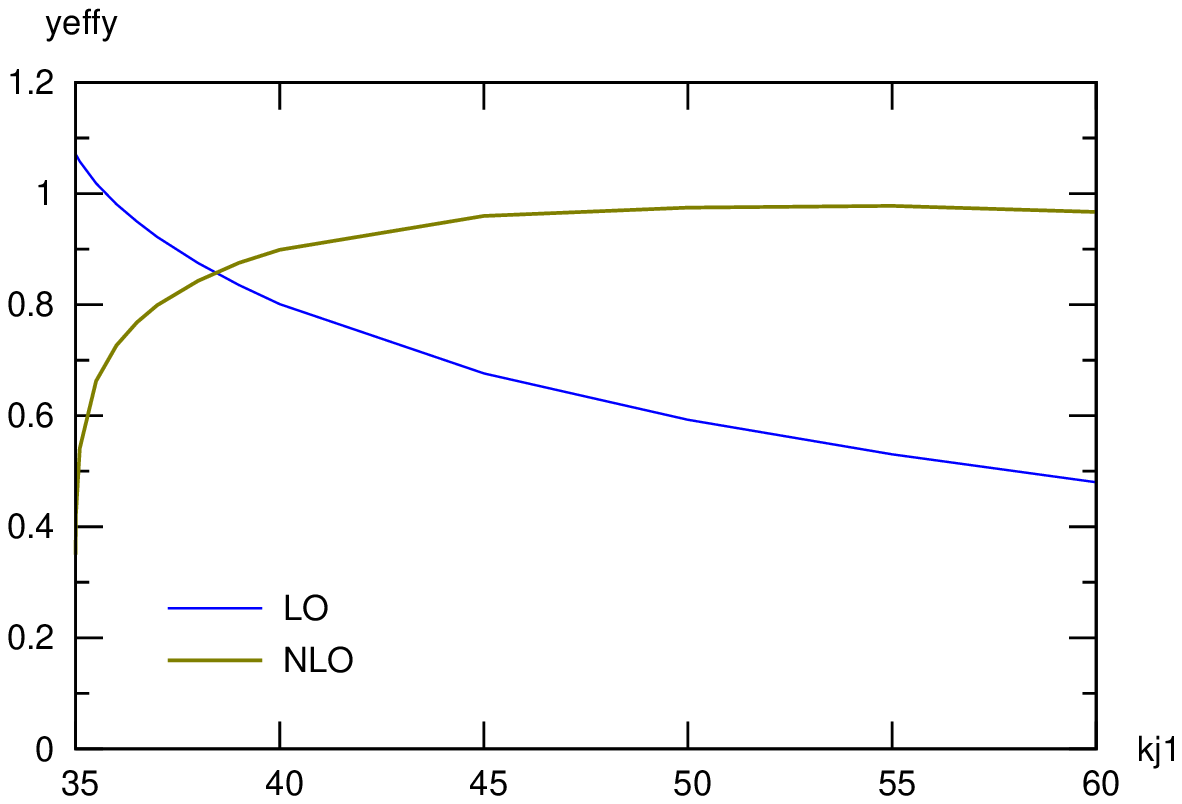}
\caption{Variation of $Y_{\rm eff}/Y$ as defined in eq.~(\protect\ref{eq:Cm_e-m_cons})
as a function of $\veckjtwo$ at fixed $\veckjone=35$ GeV for $Y=8$ and $\sqrt{s}=7$ TeV at leading logarithmic (blue) and next-to-leading logarithmic (brown) accuracy.}
\label{Fig:yeff_NLO}
\end{figure}
The value of $Y_{\rm eff}$ indicates how valid the BFKL approximation is: a value close to $Y$ means that this approximation is valid, whereas a value significantly different from $Y$ means that it is a too strong assumption in the kinematics under study.
On fig.~\ref{Fig:yeff_NLO} we show the values obtained for $Y_{\rm eff}$ as a function of $\veckjtwo$ for fixed $\veckjone=35$ GeV at a center of mass energy $\sqrt{s}=7$ TeV and for a rapidity separation $Y=8$, in the LL and NLL approximation. As found in ref.~\cite{DelDuca:1994ng}, the LL calculation strongly overestimates the cross section for transverse momenta of the jets not too close.
At NLL accuracy, the situation is much improved for
significantly different
jet transverse momenta (as needed to obtain trustable results in the fixed order approach): the effective rapidity is very close to $Y$ so that the violation of energy-momentum should be much less severe at NLL accuracy.

\section{Double parton scattering contribution to MN jets}

\begin{figure}[h]
		\centering\includegraphics[height=5cm]{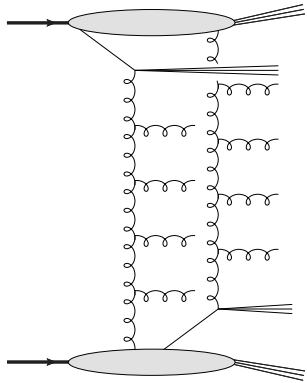}
		\caption{The DPS contribution.}
		\label{Fig:DPS}
	\end{figure}

At high energies and low transverse momenta where BFKL effects are expected to be enhanced, parton densities can become large enough that contributions where several partons from the same incoming hadron take part in the interaction could become important.
We restrict ourselves to the case of double parton scattering where there are at most two scattering subprocesses and where both these scatterings are hard, as illustrated in fig.~\ref{Fig:DPS}.
For simplicity, the order of magnitude of this contribution is evaluated at LL, which we compare with our prediction involving single parton scattering in the BFKL LL and NLL approaches. 
We use a simple factorized ansatz to compute the DPS contribution according to
	\begin{equation}
		\sigma_{\rm DPS}=\frac{\sigma_{\rm fwd} \sigma_{\rm bwd}}{\sigma_{\rm eff}} \, ,
		\label{eq:sigma_DPS}
	\end{equation}
	where $\sigma_{\rm fwd (bwd)}$ is the inclusive cross section for one jet in the forward (backward) direction and $\sigma_{\rm eff}$ is a phenomenological quantity related to the density of the proton in the transverse plane. 
	We vary $\sigma_{\rm eff}$ between 10 and 20 mb, to be consistent with
	the measurements at the Tevatron~\cite{Abe:1993rv,Abe:1997xk,Abazov:2009gc,Abazov:2014fha} and at the LHC~\cite{Aad:2013bjm,Chatrchyan:2013xxa}.
\begin{figure}[t]
	\hspace{.2cm}
	\includegraphics[width=8.5cm]{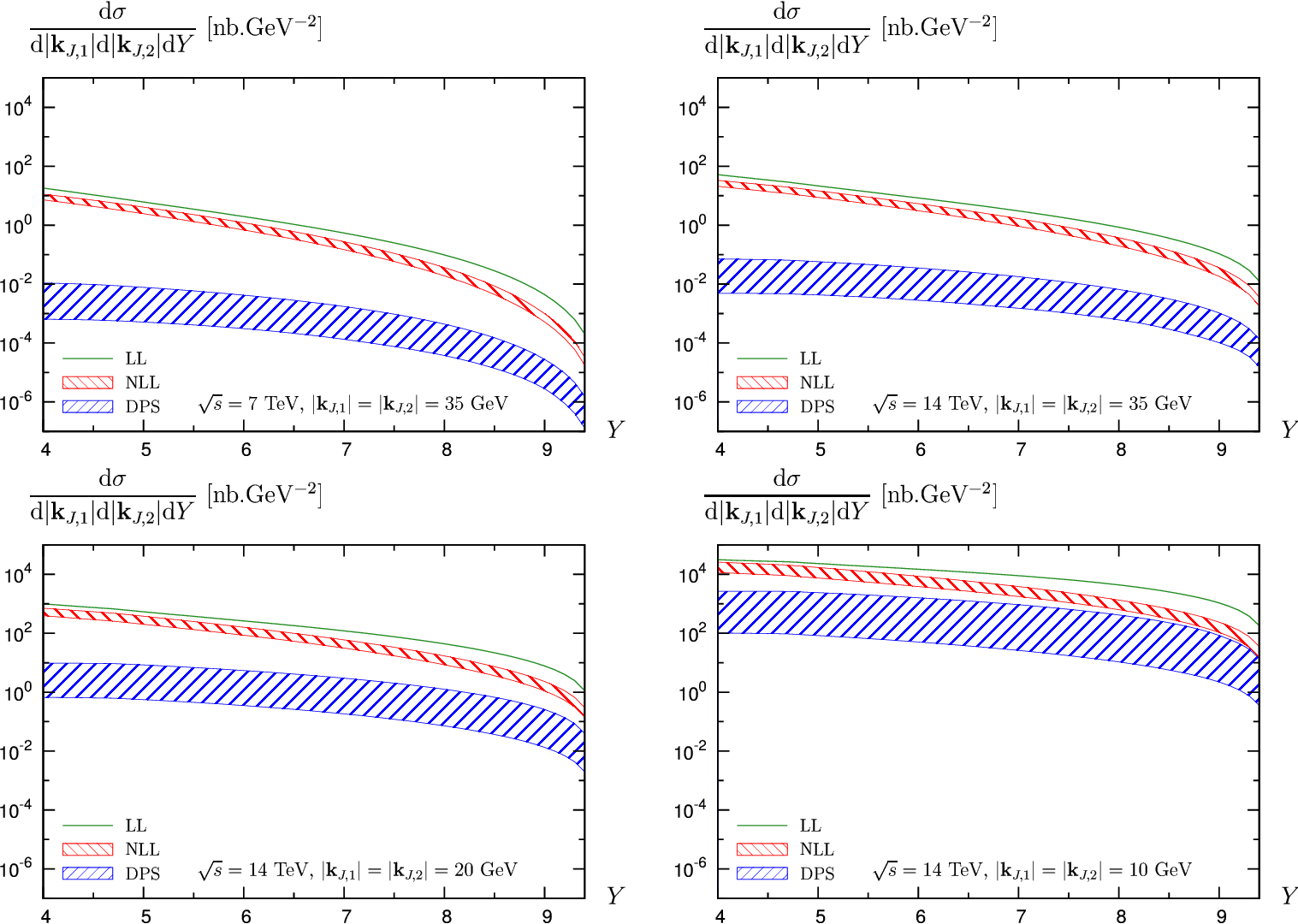}
	\caption{Comparison of the differential cross section obtained at LL (green) and NLL (red) accuracy in the BFKL approach and the DPS cross section (blue) for the four kinematical cuts described in the text.}
	\label{Fig:sigma}
\end{figure}
Each of the inclusive cross section for one jet in the forward (backward) direction is built as the convolution of the LO jet vertex with  unintegrated gluon distributions (UGD)~\cite{Kwiecinski:1997ee,GolecBiernat:1999qd,Kimber:2001sc,Hansson:2003xz,Kutak:2004ym,Jung:2004gs,Kutak:2012rf,Hautmann:2013tba}, the global normalization being fitted with CMS~\cite{Chatrchyan:2012gwa} data (see ref.~\cite{Ducloue:2015jba} for more details), for four different parametrizations.
We focus on four choices of kinematical cuts:
\begin{itemize}
	\item \kina,
	\item \kinb,
	\item \kinc,
	\item \kind.
\end{itemize}
The first choice is similar to the cuts used by the CMS analysis of azimuthal correlations of Mueller-Navelet jets at the LHC~\cite{CMS-PAS-FSQ-12-002}, as displayed in figs.~\ref{Fig:cos-cos2_blm_sym} and \ref{Fig:cos2cos-dist_blm_sym}. The other three choices correspond to the higher center of mass energy that the LHC is expected to reach soon. The last two choices correspond to lower transverse momenta at which measurements could become possible in the future, and are particularly relevant since MPI are expected to become more and more important at lower transverse momenta. The rapidities of the jets are restricted according to $0<y_{J,1}<4.7$ and $-4.7<y_{J,2}<0$. We use the MSTW 2008 parametrization~\cite{Martin:2009iq} for collinear parton densities. To estimate the 
uncertainty associated with the choice of the UGD parametrization needed to compute the DPS cross section, we use the same four parametrizations. 
Our results are displayed in fig.~\ref{Fig:sigma}.
The resulting uncertainty on the DPS cross section is rather large. Still, this cross section is always smaller than the SPS one in the LHC kinematics we considered here.
The same conclusion can be addressed 
for the impact of double parton scattering on the angular correlation between the jets.
It is only for the set of parameters giving the largest DPS contribution, {\it i.e.} at low transverse momenta and large rapidity separations, that the effect of DPS can become larger than the uncertainty on the NLL BFKL calculation.

\section{Conclusions}

The azimuthal correlations of Mueller-Navelet jets recently extracted by the CMS collaboration can be well described by a full NLL BFKL calculation supplemented by the use of the BLM procedure to fix the renormalization
scale.
We also studied two effects which are claimed to have a potential significant impact in this picture. First, we have shown that the effect of the absence of strict energy-momentum conservation in a BFKL calculation is 
expected to be tiny at NLL accuracy 
 for significantly different values of transverse momenta of the tagged jets. Second, we have shown that the order of magnitude of DPS contributions is presumably negligible for the kinematics which is under consideration at the LHC. Further studies would be required at low transverse momenta and very high center-of-mass energies. 

\section*{Acknowledgments}

B. Duclou\'{e} acknowledges support from the Academy of Finland, Project No. 273464. This work was done using computing resources from CSC -- IT Center for Science in Espoo, Finland. L.~Szymanowski was partially
supported by a French Government Scholarship.
This work is partially supported by the
French Grant ANR PARTONS No. ANR-12-MONU- 
0008-01.

\end{document}